%% file: main.v4.tex
\documentclass[%
 reprint,
superscriptaddress,
nofootinbib,
 amsmath,amssymb,
 aps, preprintnumbers,
floatfix
]{revtex4-2}

\usepackage{graphicx}% Include figure files
\usepackage{dcolumn}% Align table columns on decimal point
\usepackage{bm}% bold math
\usepackage{hyperref}% add hypertext capabilities
\usepackage[modulo]{lineno}% Enable numbering of text and display math
\usepackage[caption=false]{subfig}
\usepackage{multirow}
\usepackage{diagbox}
\usepackage{tabularx}
\usepackage[dvipsnames]{xcolor}
\usepackage{booktabs}
\usepackage{floatrow}
\usepackage{comment}

\usepackage{graphicx}
\usepackage[rightcaption]{sidecap}
\sidecaptionvpos{figure}{c}
 
\newfloatcommand{capbtabbox}{table}[][\FBwidth]

\usepackage{enumitem}

\usepackage{xspace}
\usepackage{soul}

\RequirePackage[normalem]{ulem} %DIF PREAMBLE
\RequirePackage{color}\definecolor{RED}{rgb}{1,0,0}\definecolor{BLUE}{rgb}{0,0,1} %DIF PREAMBLE
\newcommand{\unit}[1]{\ensuremath{\mathrm{\,#1}}\xspace}

\newcommand{\e}{\unit{e^{-}}}

\newcommand{\pix}{\unit{pix}}

\newcommand{\epixday}{\unit{\e /\pix/day}}
\newcolumntype{Y}{>{\centering\arraybackslash}X}

\begin{document}

\preprint{FERMILAB-PUB-24-0767-PPD, YITP-SB-2024-25}

\title{SENSEI at SNOLAB: %\texorpdfstring{1\e}{1e-} 
Single-Electron Event Rate and Implications for Dark Matter
}

\include{authors}

\date{\today}

\begin{abstract}
We present results from data acquired by the SENSEI experiment at SNOLAB after a major upgrade in May 2023, which includes deploying 16 new sensors and replacing the copper trays that house the CCDs with a new light-tight design. We observe a single-electron event rate of $(1.39 \pm 0.11) \times 10^{-5}$ \epixday, corresponding to $(39.8 \pm 3.1)$~e$^-$/gram/day. This is an order-of-magnitude improvement compared to the previous lowest single-electron rate in a silicon detector and the lowest for any photon detector in the near-infrared-ultraviolet range. We use these data to obtain a 90\% confidence level upper bound of $1.53 \times 10^{-5}$ \epixday and to set constraints on sub-GeV dark matter candidates that produce single-electron events. 
We hypothesize that the data taken at SNOLAB in the previous run, with an older tray design for the sensors, contained a larger rate of single-electron events due to light leaks. We test this hypothesis using data from the SENSEI detector located in the MINOS cavern at Fermilab. 

\end{abstract}

\maketitle

\section{\label{sec:introDM}Skipper-CCDs in dark matter searches}

Charge-coupled devices (CCDs) are pixelated silicon %-based 
detectors widely used in numerous scientific applications from imaging to particle detection~\cite{holland_fullydepleted}. Skipper-CCDs enhance their capabilities by enabling repeated, non-destructive %disruptive 
readout of the output charge, thus achieving deep sub-electron resolution~\cite{Tiffenberg:2017aac}. This feature extends the detector sensitivity to energy transfers that produce only one to a few ionization electrons, as expected from sub-GeV dark matter interactions in silicon~\cite{Essig:2011nj,Essig:2015cda,essig2024collective}.
SENSEI (\textit{Sub-Electron Noise Skipper-CCD Experimental Instrument}) is the first experiment implementing skipper-CCDs for rare-event searches, with this technology repeatedly setting leading limits~\cite{Crisler:2018gci, SENSEI:2019ibb, sensei2020, skipper_at_surface, senseicollaboration2023sensei,DAMIC-M:2023gxo} for a wide range of sub-GeV dark matter masses and models. 

The dominant backgrounds for dark matter-electron scattering have an energy of ${\cal O}$(eV) and produce single-electron (1\e) events, with two (or more) 1\e events coincident in the same or neighboring pixels mimicking 2\e ($>$2\e) events. Since the rate of 1\e background events cannot currently be modeled independently, it cannot be subtracted, and hence, the 1\e rate directly determines our sensitivity to dark matter. Reducing the 1\e background rate is crucial for improving the low-mass dark-matter search with CCD-based experiments.

Since the inception of the first skipper-CCD dark-matter search, SENSEI has been steadily reducing the 1\e rate, starting at $\sim1.14$\epixday in 2018~\cite{Crisler:2018gci}, then $(3.51\pm0.10)\times 10^{-3}$\epixday a year later~\cite{SENSEI:2019ibb}, and achieving the previous lowest recorded 1\e rate of $(1.59\pm0.16)\times10^{-4}$\epixday in 2020~\cite{sensei2020}. In 2023, the first commissioning data of the SENSEI at SNOLAB experiment produced images with a 1\e density of $(1.46\pm0.02)\times10^{-4}$\e/pix/image~\cite{senseicollaboration2023sensei} with an image exposure time of about one day. Dividing the density by the exposure time represents only an upper bound on the 1\e rate since the density contains several sources of 1\e events. In particular, in this paper, we present a measurement of both the 1\e exposure-dependent rate as well as the exposure-independent 1\e density. We use data from a new run of SENSEI science detectors at SNOLAB.

\section{\label{sec:intro1e} Single-Electron Backgrounds}

In previous work~\cite{sensei1e}, we showed that the 1\e events can be separated into two empirical components: 
an exposure-dependent rate, which contributes some number of events per pixel that scales with the exposure time of the detector, and an exposure-independent density, which does not scale with time. A dark matter signal would contribute to the exposure-dependent rate.

Exposure-dependent backgrounds might be intrinsic to the sensors, such as thermal excitations, or might have an environmental origin, such as the infrared blackbody radiation from materials around the CCDs, as well as secondary products from more energetic particles, like Cherenkov radiation or charge transfer inefficiency~\cite{sensei2020,Du_2022,Du:2023soy}. Efforts to mitigate these backgrounds include cooling the detector and the surrounding materials to temperatures between 120 and 145\,K in order to reduce dark current and blackbody radiation, a comprehensive shield design to stop radiogenic backgrounds, and analysis techniques to remove 1\e events spatially correlated with high-energy clusters.

On the other hand, the exposure-independent density is inherent to the CCD architecture and depends mainly on the operation parameters. Examples include spurious charge, produced as charge is transferred through the serial register from the pixels to the readout stage by the variation of the clock voltages, and amplifier light, emitted by the readout amplifier~\cite{janesick2001scientific,SENSEI:2019ibb,sensei1e}.

\section{\label{sec:snolab}SENSEI at SNOLAB}

We collected data for measuring the 1\e rate between November 2023 and February 2024 using the SENSEI apparatus at SNOLAB.
This is the second science run in this system; the first run was optimized for multi-electron events, and a full description of the system is given in~\cite{senseicollaboration2023sensei}. Following the first run, we installed additional CCDs of the same package design for a total of 22, and replaced all module copper trays with a new design to reduce light leaks. The skipper-CCDs are designed by Lawrence Berkeley National Laboratory and fabricated at Teledyne DALSA Semiconductor. Each has a 2.19\,g active mass of high-resistivity silicon divided into $6144 \times 1024$ pixels of $15 \times 15\,\mu \mathrm{m}^2$ and $665\,\mu$m thickness.

The larger number of copper trays inside the vessel appears to improve the inner shield. In the energy range from 500~eV to 10~keV, the background event rate is $\sim$50~events/kg/day/keV, which is a factor of 3 lower than the previous result of $\sim$140~events/kg/day/keV~\cite{senseicollaboration2023sensei}, without any dedicated effort of reducing the high-energy background.

To reduce the impact of the exposure-independent charge generated in the serial register, we performed ``hardware binning'' during readout, 
summing 32 rows at a time into the serial register so that each ``superpixel'' of the image corresponds to a $1\times32$ block of physical pixels.
Each CCD was biased with 70\,V, and we collected 300 samples per superpixel, with a resulting readout noise of 0.14\e and a single-sample readout time of 48.8\,$\mu$s. This corresponds to a readout time of about 16~minutes per image using four amplifiers, each of which reads one quadrant of the CCD (including 128$\times$4 superpixels per quadrant of overscan).
We obtained consecutive images in cycles of 0-, 2-, 6- and 20-hour exposures, designating a total of 101 images as ``commissioning'' data and 77~images as ``hidden'' data. A malfunctioning cryocooler limited the duration of a run, and we acquired data after %in 
three successive cooldowns of the system, setting the temperature of the cold finger, which cools the CCD box, to 145\,K. We ended each run when most of the cooling power was lost. 
Between cooldowns, we warmed the system to room temperature and attempted to recover the performance of the cryocooler.
Following the first cooldown (``Comm''), which established our confidence in the run parameters, we divided each of the latter two cooldowns into two hidden datasets (``Hid-1,'' ``Hid-2'') bracketed by four commissioning datasets (``Pre-1,'' ``Post-1,'' ``Pre-2,'' ``Post-2'') used to validate the system performance at the beginning and end of the hidden datasets.

\section{\label{sec:analysis}Data processing and selection}

We first apply the base data processing, calibration, and cross-talk correction described in~\cite{sensei2020,senseicollaboration2023sensei} on the commissioning and hidden data. We then use the commissioning data alone to optimize the data quality through fiducial cuts (``masks''), which we then apply directly to the hidden data to obtain an unbiased estimator of the 1\e rate and its upper bound. We follow the selection cuts presented in~\cite{senseicollaboration2023sensei} with the following modifications: at the pixel and event level, we disable the ``low-energy cluster,'' ``serial register hit,'' and ``full-well'' masks and extend the ``bleeding zone'' in the direction parallel to the serial register from 50 to 200 pixels to remove potentially very long charge trails produced by high-energy events. We compute the hot-pixel and hot-column selection with the commissioning data and use the same selection on the hidden data.

After the pixel- and event-masking procedure, we perform a binned likelihood fit of the unmasked {\it superpixel} charge histogram. We use a double Gaussian model (for the 0\e\ and 1\e\ bins) to extract the number of 1\e\ events and the noise. We use this information to design a ``noisy-image'' mask, where we remove an entire image if the $p$-value of the fit is below 0.005 or the noise in unmasked superpixels differs from the calibration by more than 10\%. 

In addition, we implement a ``hot image" mask separately for the commissioning and the hidden data. For commissioning (hidden) data, we mark a quadrant in a given image as hot if its number of 1\e events is more than 3$\sigma$ away from the mean of all images with the same exposure in the processed commissioning (hidden) data, normalized to the number of unmasked superpixels. 

We designate the quadrant with the lowest 1\e rate in the commissioning data as the ``Golden'' Quadrant. We observe that the 1\e rates are higher in the Post-1 and Post-2 data than in the Pre-1 and Pre-2 data, which suggests that the ends of the Hid-1 and Hid-2 data might have similarly elevated 1\e rates due to the cryocooler malfunctioning. Since we do not have a direct measurement of the CCD temperature, we choose two ``Witness'' Quadrants, whose rate variations in the commissioning data follow the same trend as the Golden Quadrant. %in the commissioning data.  
We first unblind these Witness Quadrants and measure their 1\e rates to identify periods where the data quality is good and the cryocooler is performing well while keeping the Golden Quadrant hidden. 
We then proceed to unblind the Golden Quadrant. 

\section{\label{sec:results}Results}

\begin{figure}

  \includegraphics[width=1\linewidth]{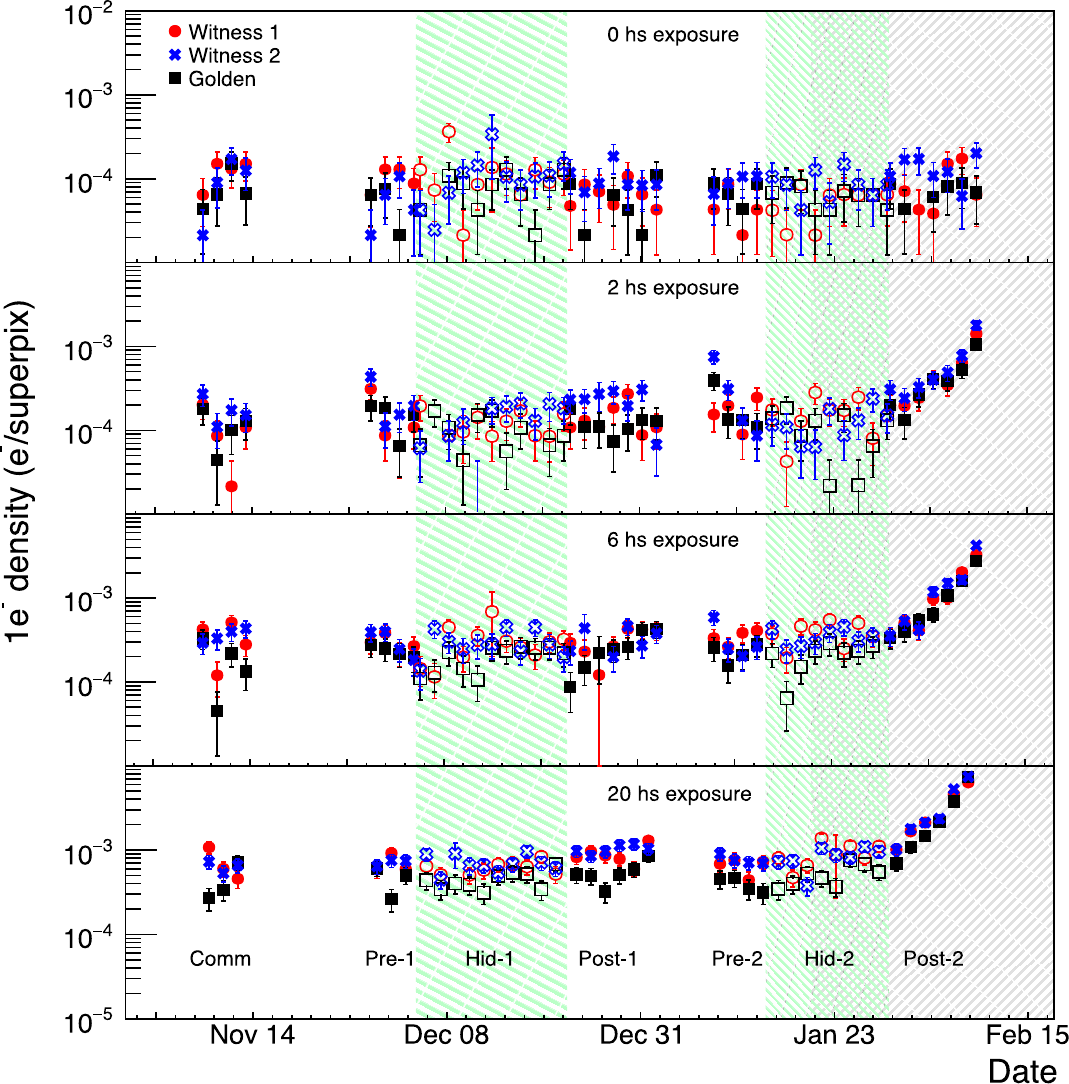}
  \caption{The 1\e density per superpixel per image (after applying the pixel masking) versus the acquisition start time for the 0-hour (top), 2-hour (middle-top), 6-hour (middle-bottom), and 20-hour (bottom) exposures. Red circles, blue crosses, and black squares are the data for Witness Quadrant 1, Witness Quadrant 2, and the Golden Quadrant, respectively.   Green shading and open markers indicate time periods during which we took hidden data, while solid markers indicate commissioning-data periods.  
  Gray shading indicates data removed from the analysis due to the malfunctioning cryocooler. For the time periods without data points the system was warm and the data acquisition was off.}
  \label{fig:rates}
\end{figure}

The hot-image mask flags one image as hot in Witness Quadrant 1. 
After removing this image and applying all selection cuts, between 80\% and  95\% of the pixels survive the masking (depending on the exposure). In Fig.~\ref{fig:rates}, we present the 1\e density per superpixel per image (after applying the pixel masking) versus the acquisition start time for the various exposures.

\begin{figure}

  \includegraphics[ trim={0.0cm 0.0cm 0.0cm 0.0cm},clip,width=0.9\linewidth]{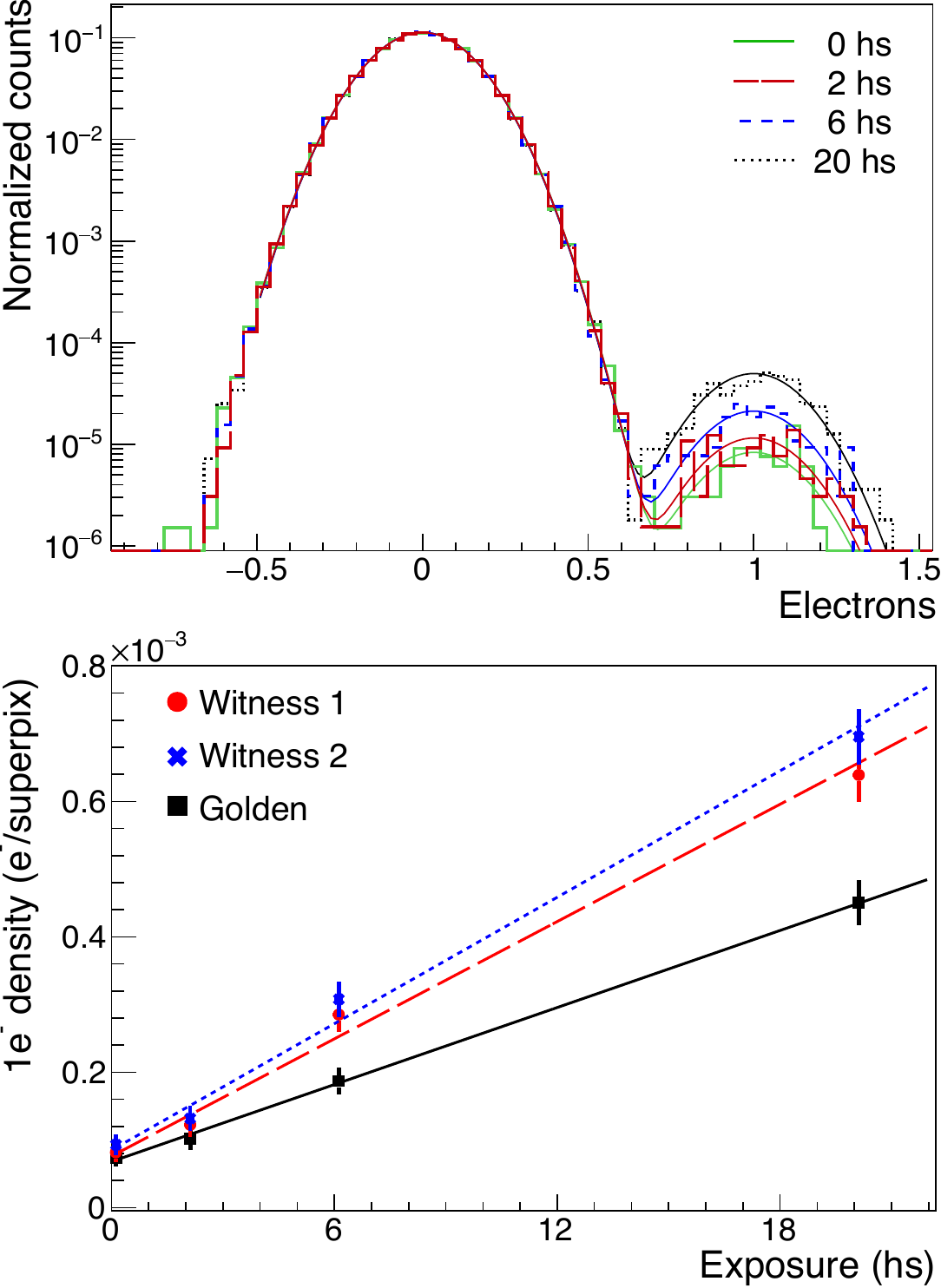}%12.2cm
 
  \caption{\textbf{Top:} Unmasked superpixel charge distribution for 0- (solid-green), 2- (long-dashed-red), 6- (short-dashed-blue), and 20-hour (black-dotted) exposures, including all images of the Golden Quadrant. Fits correspond to a double Gaussian peak from which the 1\e density is extracted. \textbf{Bottom:} 1\e density per superpixel per image as a function of the exposure. We extract the exposure-dependent rates (slope) and exposure-independent densities (intercept) with a linear fit. We summarize the results of these fits in Table~\ref{table:rates}. We show results for the two Witness Quadrants (red circles with dashed line and blue crosses with dotted line) and the Golden Quadrant (black squares with solid line).}
  \label{fig:fits}
\end{figure}

We create histograms of the unmasked superpixel charge for the various exposures %by exposure 
and obtain the number of 1\e events per superpixel using a double Gaussian function that fit the 0\e and 1\e peaks. 
The top panel of Fig.~\ref{fig:fits} shows the histograms for the Golden Quadrant for the entire hidden data (Hid-1 and Hid-2) outside the gray-shaded time periods, normalized by the number of entries, for the 0-, 2-, 6-, and 20-hour exposures, along with their respective double-Gaussian fits. The number of 1\e events increases for larger exposures, although most superpixels remain empty. We present the fitted 1\e density versus the exposure in the bottom panel of Fig.~\ref{fig:fits} for the two Witness and Golden Quadrants. We also present the linear fits from which the exposure-dependent (slope) and exposure-independent (intercept) values are extracted. We then convert the results from superpixel to pixel, calculate the 90\% confidence level upper bound for the exposure-dependent rate, and summarize the results in Table~\ref{table:rates}. We note that the readout time is included in the exposure calculation.

\begin{nolinenumbers}
\begin{table*}
\centering
\renewcommand{\arraystretch}{1.25}
\begin{tabularx}{\textwidth}{l | Y  Y | Y Y} 
\toprule
%\multicolumn{1}{c}{} & \multicolumn{2}{c}{\textbf{ superpixel}} & \multicolumn{2}{c}{\textbf{pixel}} \\
%\cmidrule(rl){2-3} \cmidrule(rl){4-5}
Quadrant  & Exposure independent  &  Exposure dependent  &  Exposure dependent  & 90\% U.L.\\
      & $\times 10^{-5}$\e/superpix/image  &  $\times 10^{-4}$\e/superpix/day  &  $\times 10^{-5}$\epixday & $\times 10^{-5}$\epixday \\
%\multicolumn{1}{c|}{} & \multicolumn{1}{c}{$\times 10^{-5}$\e/superpix/image} & \multicolumn{1}{c|}{$\times 10^{-4}$\e/superpix/day} & \multicolumn{2}{c}{$\times 10^{-5}$\epixday}\\
\midrule
Golden     &  $6.94 \pm 0.85$  &  $4.44 \pm 0.35$  &  $1.39 \pm 0.11$  &  $1.53$ \\
Witness 1~~  &  $7.64 \pm 0.97$  &  $6.82 \pm 0.43$  &  $2.13 \pm 0.13$  &  $2.30$ \\
Witness 2  &  $8.70 \pm 1.03$  &  $7.13 \pm 0.45$  &  $2.23 \pm 0.14$  &  $2.41$ \\
\bottomrule
\end{tabularx}
\caption{The exposure-independent 1\e density and the exposure-dependent 1\e rate per superpixel extracted from the bottom panel in Fig.~\ref{fig:fits}. The two columns on the right provide the exposure-dependent per-pixel 1\e\ rates and the corresponding 90\% confidence level upper limit for the Golden, Witness 1, and Witness 2 Quadrants.  %and converting to pixel. 
%We also present the 90\% confidence level upper limit for the given exposure-dependent results.
}
\label{table:rates}
\end{table*}
\end{nolinenumbers}

For the Golden Quadrant, we obtain an exposure-dependent rate of $(1.39 \pm 0.11) \times 10^{-5}$\,\epixday and a 90\% confidence level upper limit of $1.53 \times 10^{-5}$~\epixday, the lowest ever achieved with a silicon detector and an order of magnitude improvement over the previous best-published values in~\cite{sensei2020,senseicollaboration2023sensei}. We note that the rates in the Witness Quadrants also improve on the previously best published values.

Although a quantitative analysis is beyond the scope of this paper, we note that Cherenkov radiation is unlikely to be the main source of the remaining exposure-dependent 1\e rate~\cite{Du_2022,Du:2023soy}, since the rate of high-energy events that can cause such radiation is small (see SM). 
Instead, the exposure-dependent 1\e rate can either be produced by detector dark counts or by an external source of low-energy radiation such as infrared photons or dark matter.  
We can interpret the measured upper limit on the 1\e-rate as an upper limit on the various 1\e background sources or as an upper limit on possible dark-matter-induced 1\e events.  

During the first run at SNOLAB~\cite{senseicollaboration2023sensei}, we did not separate the exposure-dependent and exposure-independent contributions to the 1\e events. Nevertheless, the lowest 1\e density obtained was $\sim 1.44 \times 10^{-4}$ \e/pix/image with an exposure close to one day, which is larger than expected based on this new measurement (assuming the exposure-independent density does not change with the binning factor, the mean expected 1\e density in the Golden Quadrant would be $8.5\times 10^{-5}$\e/pix/image). 
We hypothesize, although we cannot conclusively prove, that the 1\e density in the first run had a significant contribution from blackbody radiation leaking onto the CCD, which was reduced with a new tray design that has fewer light leaks. 

We next present evidence for the light-leak hypothesis as an explanation of the larger number of 1\e events in the first run, using data acquired with the SENSEI detector near the MINOS cavern.

\section{\label{sec:minos} Comparison with SENSEI at MINOS and discussion}

A contribution from blackbody radiation leaking onto the CCD may explain the higher 1\e rate in the first SNOLAB run~\cite{senseicollaboration2023sensei}. With the SENSEI setup at MINOS, we previously identified an elevated 1\e rate in CCD regions with a direct line of sight to warm surfaces~\cite{sensei2020}. The SENSEI setup at SNOLAB has the CCD trays enclosed in a cold copper box, but the box has openings to room-temperature parts inside the vessel. Openings in the CCD trays used in the first run could allow blackbody radiation to reach the CCDs and source 1\e events. These would then be reduced by the improved tray design used in the second run.

The tray covers in the first run had large openings at the two leaf springs and in the two corners opposite the flex cable. In addition, the tray base and cover sandwich the CCD flex cable so the cover and base edges do not touch. This results in a thin gap at the long edges of the tray, where there is no flex cable. The front surfaces of the CCDs are fully exposed to any light that enters the tray, while the back surfaces are covered by a silicon substrate and the copper tray base. We redesigned the trays for the second run with closed corners, but the gaps at the leaf springs and edges remain (see SM for pictures of the trays). %~\ref{fig:trays} in supplemental materials for pictures of the trays).

To further explore the light-leak hypothesis, we ran a series of tests at the MINOS setup 
%using a slightly-modified MINOS setup from the one 
described in~\cite{sensei2020}, where the elements inside the vessel surrounding the copper tray are warm. 
Later efforts on that setup allowed us to reduce the exposure-dependent 1\e rate to $\sim1\times 10^{-4}$\epixday, not previously reported. We subsequently installed two skipper-CCDs of the same type and packaging design as those used in SNOLAB. We also improved the lead shield outside the vacuum vessel. The background event rate in the 500~eV to 10~keV energy region is measured to be $\sim$600~events/kg/day/keV, which is about 6 times smaller than in the previous MINOS run~\cite{sensei2020}.

\begin{figure}
  
  \centering\includegraphics[width=1.0\linewidth]{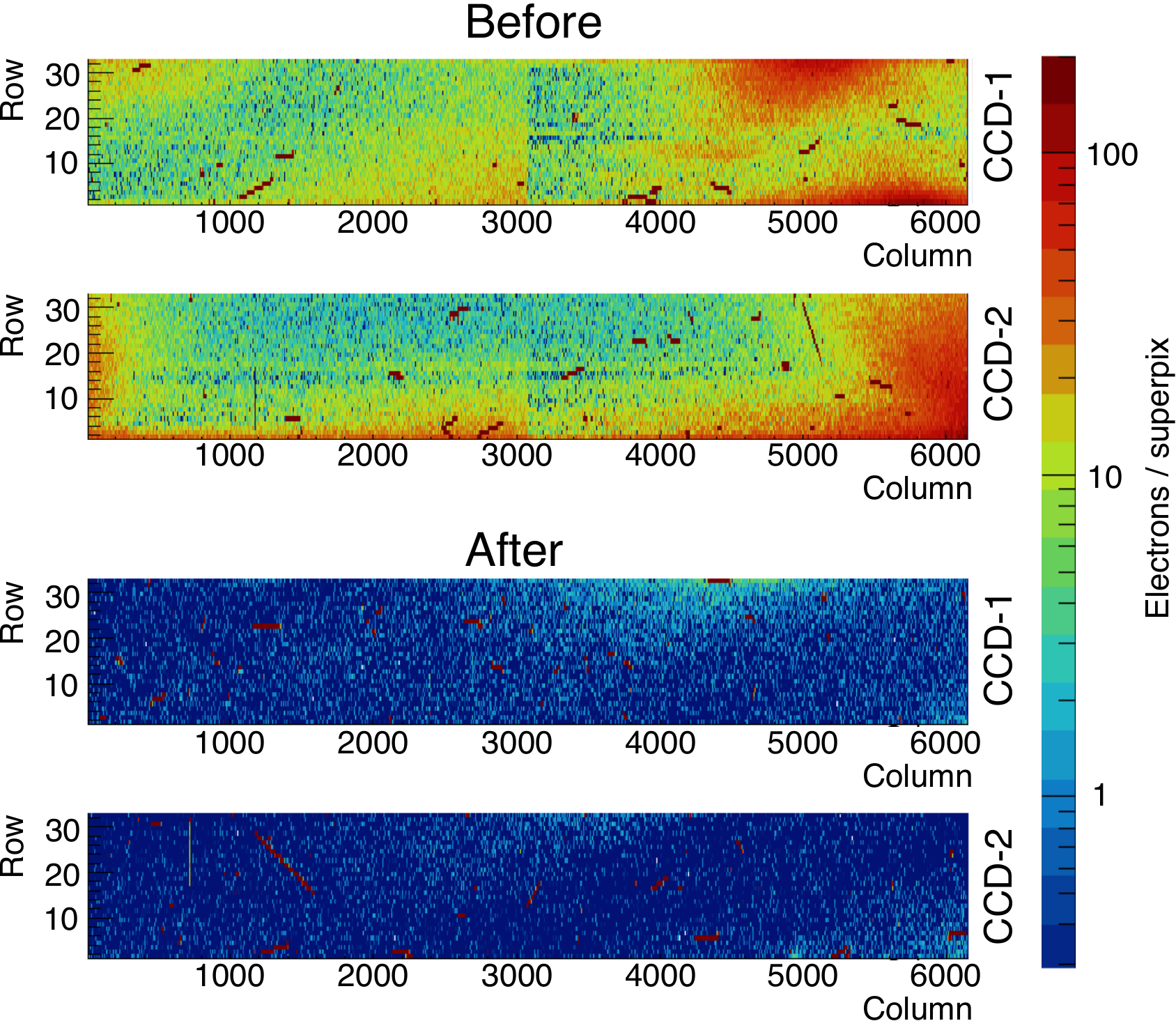}
  \caption{CCD images obtained with an LED turned on inside the MINOS vessel before (top) and after (bottom) replacing the copper trays that hold the CCDs with a newer design, which has fewer light leaks.  We display the CCDs and quadrants matching their physical position in the copper tray. The color scale indicates the number of electrons in a superpixel. Saturated tracks correspond to high-energy events, such as muons.}
  
  \label{fig:LED}
\end{figure}

%At the beginning of this study, the 
The skipper-CCDs were first housed in a tray with open corners identical to those used in the first SNOLAB run but with copper tape sealing the leaf springs and corners. The 1\e rate was about $8 \times 10^{-5}$\epixday. We placed an LED in the MINOS vessel's vacuum pump line to illuminate the CCD tray from the flex-cable end. The upper half of Fig.~\ref{fig:LED} shows an image with 6~hour exposure and similar operation parameters as in~\S\ref{sec:snolab}, with the LED on. We observed a pattern consistent with light leaking through the edges of the CCD tray.

To reduce this apparent light leak, we replaced the copper trays with those used in the second SNOLAB run (i.e., identical to those used for the results reported in this paper). These have closed corners, and in addition, we sealed the leaf springs, corners, and most of the edges with copper tape. The lower half of Fig.~\ref{fig:LED} shows an image with the same LED power and operating parameters as the upper half, and we observe that the light leaks are much reduced.

To further understand the impact of the light leaks on the 1\e rate, we acquired data with identical parameters as used in SNOLAB and with 0-, 2-, and 6-hour exposures (omitting the 20-hour exposures due to the higher background from cosmic-ray muons). 
We run the same analysis pipeline as the one used on the SNOLAB data (this was not a hidden analysis, but no analysis choices were made based on these data). We summarize the results in Table~\ref{table:MINOSrates}. The lowest dark current obtained is $3.43 \times 10^{-5}$\epixday, more than a factor two improvement compared to that before the tray change, which strongly supports the hypothesis that light leaks, even through an indirect path, are an important 1\e background. While this does not conclusively 
prove that light leaks are the source of the higher 1\e rates obtained during the first SNOLAB run, it does provide compelling evidence. 

\begin{nolinenumbers}
\begin{table}
\centering
\renewcommand{\arraystretch}{1.25}

\begin{tabularx}{\textwidth}{Y | Y  Y} 
\toprule

\multirow{2}{*}{CCD-1} & $3.49 \pm 0.13$  & $4.49 \pm 0.15$ \\ 
     & $7.58 \pm 0.21$   & $7.57 \pm 0.21$ \\
     \midrule
\multirow{2}{*}{CCD-2} & $8.19 \pm 0.25$   & $4.36 \pm 0.15$ \\ 
       & $6.88 \pm 0.19$  & $8.23 \pm 0.22$ \\

\bottomrule
\end{tabularx}
\caption{
Results for exposure-dependent rate for all MINOS quadrants obtained after the tray replacement. Result are in $\times 10^{-5}$\epixday and arranged according to the physical position of the four quadrants. The best 1\e rate obtained before the intervention is $8 \times 10^{-5}$\epixday. See Table~\ref{table:MINOSratesSup} in supplemental materials for the exposure-independent 1\e densities.}
\label{table:MINOSrates}
\end{table}
\end{nolinenumbers}

\begin{figure*}[t]
    \centering
    \includegraphics[width=0.4\textwidth]{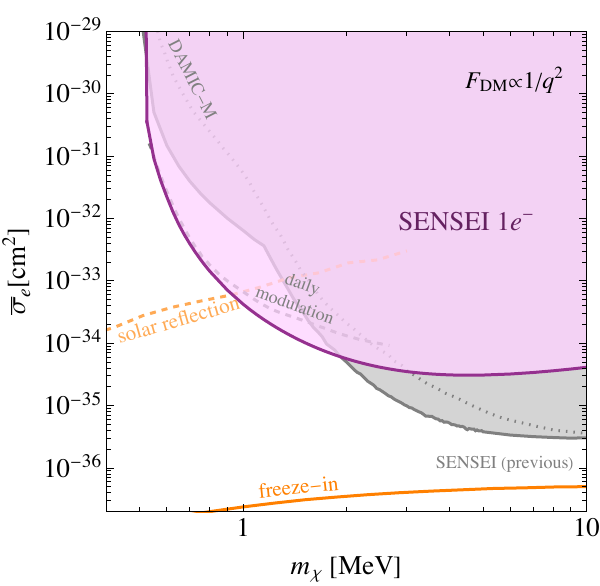}
   %\hspace*{-0.1in}
    \includegraphics[width=0.4\textwidth]{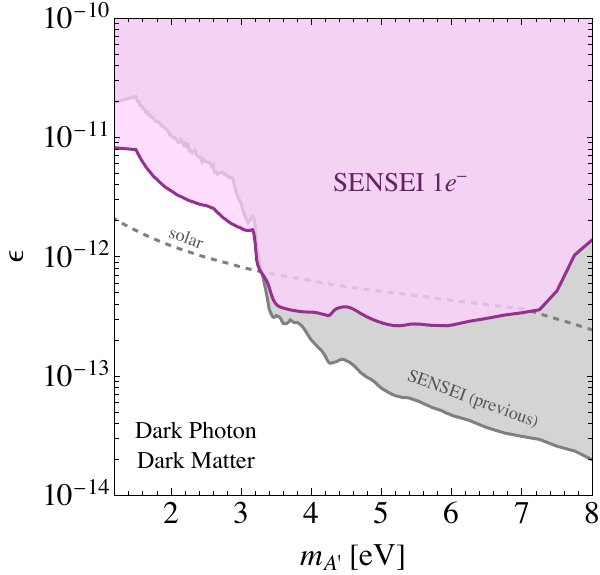}
    \caption{\textbf{Left:} Solid pink line and shade correspond to the 90\%~C.L.~constraints on dark-matter-electron cross section, $\overline{\sigma}_e$, 
versus dark-matter mass, $m_\chi$, for light mediators using the 1\e channel. The dashed-orange line represents the constraint for the solar-reflected halo dark-matter, assuming a dark-photon mediator~\cite{Emken:2024nox,An:2021qdl}. The orange solid line is the freeze-in benchmark target~\cite{Essig:2011nj,Essig:2015cda,Chu:2011be,Dvorkin:2019zdi}. The gray shade, solid line, corresponds to previous SENSEI constraints using hidden analyses~\cite{sensei2020,senseicollaboration2023sensei}, the dotted line is from a non-hidden DAMIC-M analysis~\cite{DAMIC-M:2023gxo}, and the dashed line is from a daily modulation search from DAMIC-M assuming an ultralight dark-photon mediator~\cite{DAMIC-M:2023hgj}. \textbf{Right:} Bounds on the kinetic-mixing parameter, $\epsilon$, versus the dark-photon mass, $m_{A'}$, for dark-photon dark matter absorption, including solar bounds in a dashed-gray line~\cite{Bloch:2016sjj,An:2013yfc,Redondo:2013lna}
\label{fig:dmlimit}}
\end{figure*}

\section{\label{sec:impact}Impact on dark-matter sensitivity}

Using the 1\e rate from the Golden Quadrant in Table~\ref{table:rates}, we present in Fig.~\ref{fig:dmlimit} the 90\% confidence level limit on halo dark-matter-electron scattering via a light mediator~\cite{Essig:2011nj} (left panel), and for the absorption of dark-photon dark matter~\cite{DIMOPOULOS1986145,PhysRevD.35.2752,An:2014twa,Bloch:2016sjj,Hochberg:2016sqx} (right panel). Solid pink lines and shading represent the results from this work, while gray lines and shading represent previous SENSEI constraints~\cite{sensei2020,senseicollaboration2023sensei}. The scattering rates were calculated using {\tt QEDark}~\cite{Essig:2015cda,QEdark} to facilitate comparison with other bounds in the literature, and the halo parameters in~\cite{Baxter:2021pqo}\footnote{$v_{\rm esc}$=544 km/s, $v_0$=238 km/s, $v_E$=250 km/s, $\rho_{\rm DM}$=0.3 GeV/cm$^3$}. We implement the ionization model in~\cite{Ramanathan_2020} to estimate the number of secondary electron-hole pairs produced by the recoiling electron. 
These constraints significantly improve on previous limits for low dark matter masses. We present in the SM bounds on dark matter scattering off electrons via a heavy mediator with {\tt QEDark}, as well as constraints using {\tt QCDark}~\cite{Dreyer:2023ovn,QCDark}.

In summary, we presented a measurement of the 1\e rate with skipper-CCDs at SNOLAB, and we obtained an order-of-magnitude improvement compared to the previous rates %World record 
in silicon detectors and the lowest rate for any photon detector in the near-infrared-ultraviolet range. We discussed the impact on the 1\e rate of light leaks from blackbody radiation and presented an improved %and world-leading 
constraint on dark matter-electron interactions through light-mediator scattering and absorption. The next steps in SENSEI consist of further pushing these boundaries with an improved package design and cleaner shield, as well as with new analysis techniques. 

\begin{acknowledgments}
We thank Andrew Lathrop and our beloved cryocooler (who gave it all until the very end). 
We are grateful for the support of the Heising-Simons Foundation under Grant No.~79921.
This work was supported by Fermilab under U. S.~Department of Energy (DOE) Contract No.~DE-AC02-07CH11359. 
The CCD development work was supported in part by the Director, Office of Science, of the DOE under No.~DE-AC02-05CH11231. RE acknowledges support from DOE Grants DE-SC0017938 and DE-SC0025309, and Simons Investigator in Physics Awards~623940 and MPS-SIP-00010469. 
We would like to thank SNOLAB and its staff for support through underground space, logistical and technical services. SNOLAB operations are supported by the Canadian Foundation for Innovation and the Province of Ontario, with underground access provided by Vale at the Creighton mine site.
TV is supported, in part, by the Israel Science Foundation (grant No. 1862/21), %by the Binational Science Foundation (grant No.\ 2020220), 
by the NSF-BSF (grant No.\ 2021780) and by the European Research Council (ERC) under the EU Horizon 2020 Programme (ERC-CoG-2015, Proposal No.\ 682676 LDMThExp).
RE and TV acknowledge support from the Binational Science Foundation (grant No.\ 2020220). 
IB is grateful for the support of the Alexander Zaks Scholarship, The Buchmann Scholarship, and the Azrieli Foundation.
This manuscript has been authored by Fermi Research Alliance, LLC under Contract No. DE-AC02-07CH11359 with the U.S.~Department of Energy, Office of Science, Office of High Energy Physics. The United States Government retains and the publisher, by accepting the article for publication, acknowledges that the United States Government retains a non-exclusive, paid-up, irrevocable, world-wide license to publish or reproduce the published form of this manuscript, or allow others to do so, for United States Government purposes.
\end{acknowledgments}

\newpage 
%\vskip 6cm %\clearpage 

\begin{center}
    \large{\textbf{Supplemental Materials}}
\end{center}
\setcounter{section}{0}

\section{Design of skipper-CCD trays}\label{app:trays}

\begin{figure}[h]
  
  \centering\includegraphics[width=1.0\linewidth]{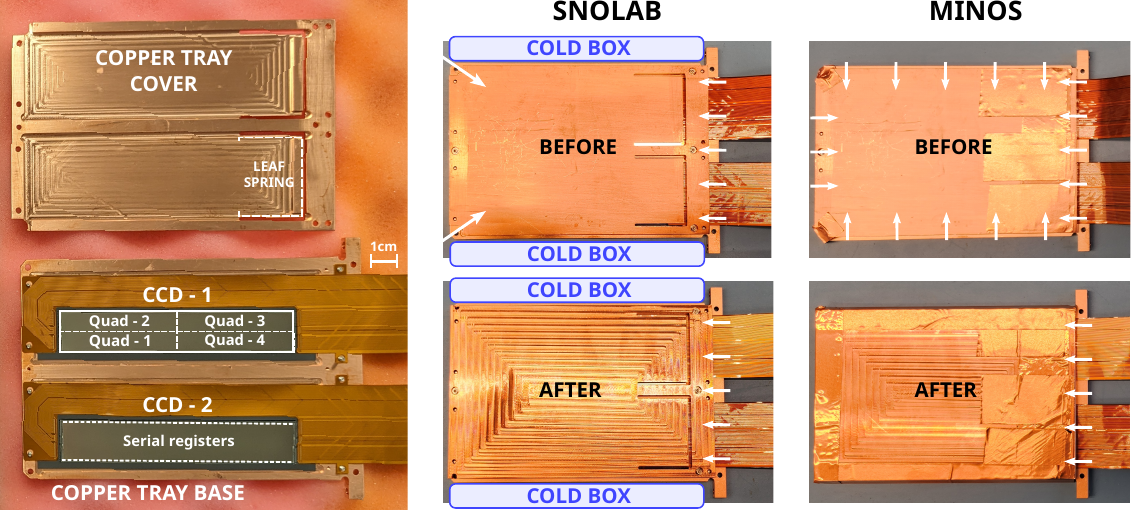}
  \caption{
  %Copper tray design at SNOLAB and MINOS before and after replacing the trays. 
  \textbf{Left:} The cover and base of the older tray design, revealing the position of two CCDs inside. 
\textbf{Middle:}  Copper trays in SNOLAB for the run 1 (top) and run 2 (bottom).  We use the bottom tray design to take the 1\e rate data reported in in the main paper.
White arrows indicate the location where we expect light leaks. 
\textbf{Right:} Copper tray configuration for the MINOS light-leak studies described in \S\ref{sec:minos} of the main paper.  The top plot shows the copper tray design with open corners, while the bottom one has the corners closed with copper tape.}  
  \label{fig:trays}
\end{figure}

We used new trays for holding the skipper-CCDs for the second SENSEI run at SNOLAB, which led to the  1\e rate measurement reported in this paper. The left panel of Fig.~\ref{fig:trays} shows the cover and base of the older tray design. The old tray cover has large openings at the two leaf springs and in the two corners opposite the flex cable. The new trays have closed corners, although the gaps at the leaf springs and edges remain. 

The center two photos in Fig.~\ref{fig:trays} show the tray configuration in the SNOLAB system, with the top and bottom photos showing the old and new tray designs, i.e., for run 1 and run 2, respectively. We indicate with white arrows where we expect light leaks. The long edges of the tray fit into the slots in the cold box and are effectively sealed. The leaf-spring openings in both tray designs are uncovered but face other trays that shield these openings. We then expect the open corners in the old design to be the dominant light leak for the first SNOLAB run, so the new tray design significantly improves the light tightness. We also expect to have some minor light leaks on the flex cable side.

The right two photos in Fig.~\ref{fig:trays} show the tray configuration for the MINOS tests described in \S\ref{sec:minos} of the main paper. We performed the first test with the LED using the tray design of run 1 at SNOLAB but with copper tape covering the corners and leaf spring. We thus expect to have most of the light leaks through the tray edges. For the second test, we used the tray design in the second SNOLAB run, with copper tape on the springs and edges. In this case, we expect some light leaks to remain on the side of the flex cables, which we were unable to seal with copper tape. 

\section{High-Energy Event Spectrum}\label{app:HE-spectrum}

\begin{figure}[t]
  
  \centering\includegraphics[width=1.0\linewidth]{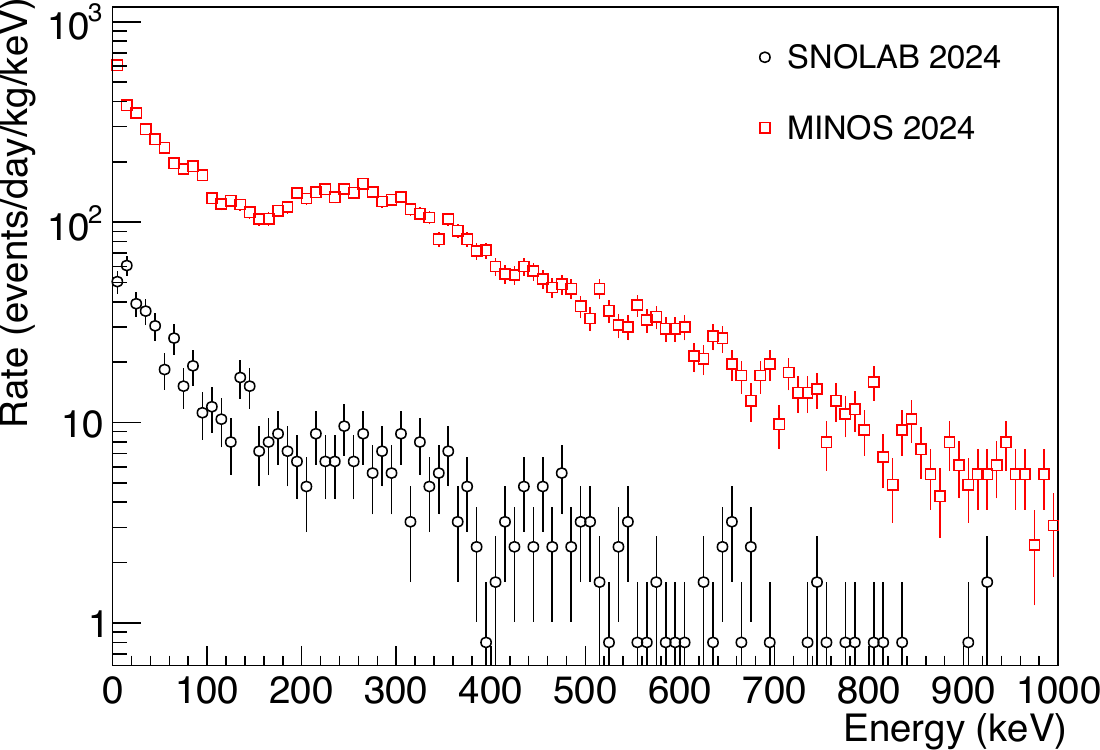}
  \caption{High-energy event spectrum from 500~eV to 1~MeV. A subset of masks have been applied to remove instrumental backgrounds. Black circles show the spectrum from SNOLAB using a selection of good quadrants and images from the commissioning data (we exclude the parts of Post-2 for which the system was warm). The red squares show the spectrum from MINOS after the tray replacement.}
  
  \label{fig:DRU}
\end{figure}

In Fig.~\ref{fig:DRU}, we present the high-energy event spectrum ranging from 500~eV to 1~MeV. Within the 500~eV to 10~keV energy range, the background event rate at SNOLAB (black circles) is approximately 50~events/kg/day/keV, which is roughly 3 times lower than that observed during the first SENSEI run~\cite{senseicollaboration2023sensei}. The event rate at MINOS (red squares) is approximately 600~events/kg/day/keV, 5 times lower than the previous MINOS run~\cite{sensei2020}. Notably, we did not conduct a surface etch on our copper components, a procedure that could potentially further reduce these background rates.

\section{MINOS exposure-independent \texorpdfstring{1\e}{1e-} rate}\label{app:minosExpIndep}

Table~\ref{table:MINOSratesSup} shows the exposure-independent 1\e\ densities obtained in the two CCDs in the SENSEI MINOS setup. 

\begin{nolinenumbers}
\begin{table}[ht]
\centering
\renewcommand{\arraystretch}{1.25}

\begin{tabularx}{\textwidth}{Y | Y  Y} 
\toprule

\multirow{2}{*}{CCD-1} & $4.99 \pm 0.43$   & $6.03 \pm 0.52$  \\ 
     &  $6.24 \pm 0.74$   & $6.56 \pm 0.75$ \\
     \midrule
\multirow{2}{*}{CCD-2} & $ 12.23 \pm 0.88$     &$ 9.94 \pm 0.54$  \\ 
       & $7.53 \pm 0.70$    & $ 6.52 \pm 0.80$\\

\bottomrule
\end{tabularx}
\caption{Results for the exposure-independent densities for all MINOS quadrants obtained after replacing the trays with the closed-corner design. Result are in $\times 10^{-5}$\e/superpix/image and arranged according to the physical position of the quadrants. See Table~\ref{table:MINOSrates} in \S\ref{sec:results} for the exposure-dependent 1\e\ rates.}
\label{table:MINOSratesSup}
\end{table}
\end{nolinenumbers}

\section{Additional Dark Matter Constraints}\label{app:results}

\begin{figure}
    \centering
    \includegraphics[width=0.7\linewidth]{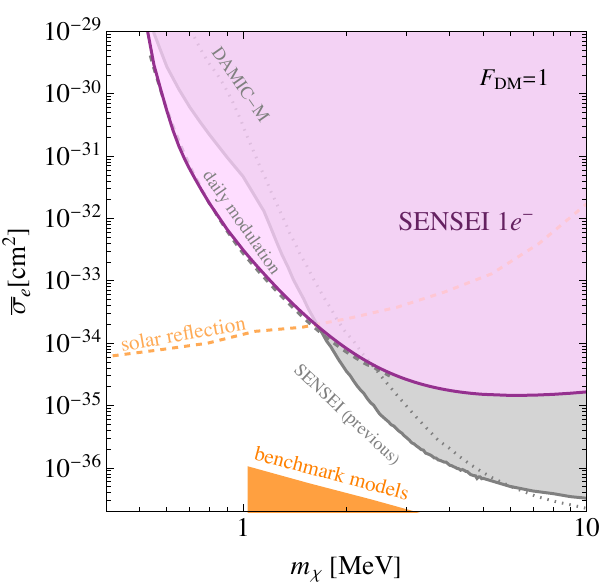}
    \includegraphics[width=0.7\textwidth]{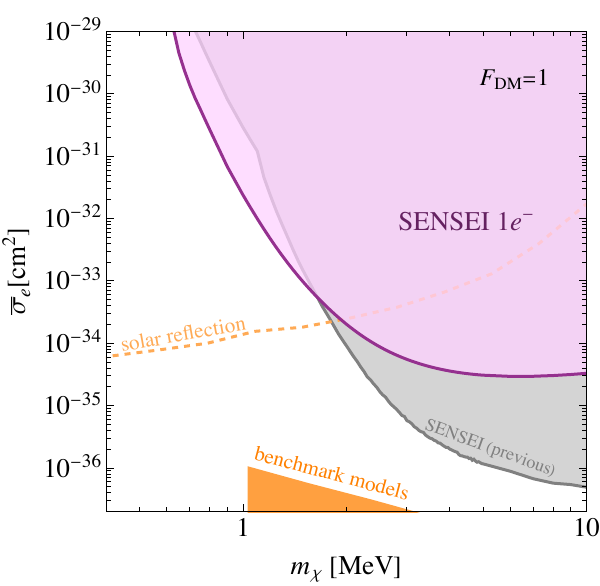}
    \includegraphics[width=0.7\linewidth]{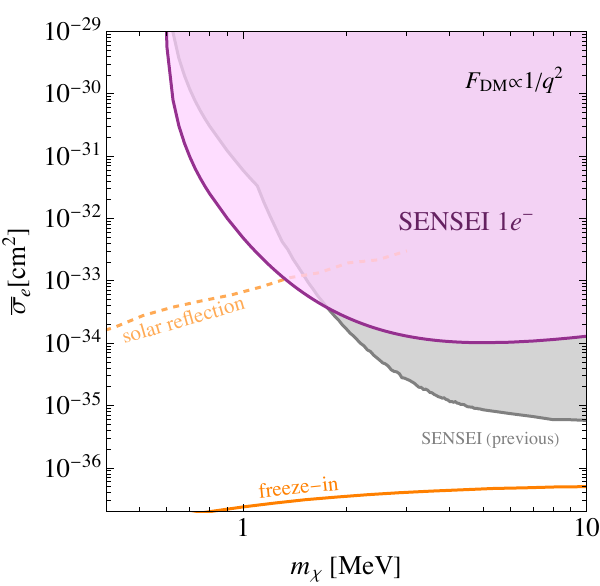}
    \caption{
    90\%~C.L.~upper-limits on the dark matter-electron scattering cross-section for heavy mediator calculated with {\tt QEDark} without including screening effects  (\textbf{top}) and {\tt QCDark} (\textbf{middle}); and for a light mediator, also with {\tt QCDark} (\textbf{bottom}). We show results from this work in pink lines and shades, while gray solid lines and shaded regions show SENSEI results using hidden analyses~\cite{sensei2020,senseicollaboration2023sensei}. Gray dotted line is from DAMIC-M (using a non-hidden analysis)~\cite{DAMIC-M:2023gxo}, the dashed-line represents a daily modulation limit from DAMIC-M assuming a heavy dark photon mediator~\cite{DAMIC-M:2023hgj}, and dashed-orange lines are the solar-reflected halo dark-matter bounds assuming a dark photon mediator~\cite{Emken:2024nox,An:2017ojc,An:2021qdl}. Orange shades correspond to benchmark targets from~\cite{Essig:2011nj,Essig:2015cda,Essig:2022dfa,Boehm:2003hm, Lin:2011gj,Izaguirre:2015yja,Hochberg:2014dra,Kuflik:2017iqs,DAgnolo:2019zkf,Chu:2011be,Dvorkin:2019zdi}.
\label{fig:QEdark} }
\end{figure}

The top panel of Fig.~\ref{fig:QEdark} shows the constraints on dark matter-electron scattering through a heavy mediator calculated with {\tt QEDark}~\cite{Essig:2015cda,QEdark} without screening to facilitate comparison with existing bounds in the literature. %Note that in the {\tt QEDark} figure, the SENSEI@SNOLAB results are from all the data, not only those using the hidden data, to facilitate comparison with other unblinded searches. 
In addition, we show in the middle and bottom panels of Fig.~\ref{fig:QEdark} constraints for heavy and light mediator, calculated with {\tt QCDark}~\cite{Dreyer:2023ovn,QCDark}. 

\section{Data release}\label{app:opendata}

The repository \url{https://github.com/sensei-skipper/DataReleases} contains the images corresponding to the hidden dataset for the Golden Quadrant used to produce the results in this work. We release the images stored in ROOT files separated by exposure time (in seconds). The ROOT tree inside each file contains the pixel information, including position in the image, charge after calibration, image identification number, and which masks were applied. We also include a ROOT macro to obtain the 1\e density for each exposure and fit the 1\e rate. We plot the dark matter constraints in this paper using the 90\%~C.L. of this rate.

\newpage
\bibliographystyle{apsrev4-1}

\bibliography{sensei}

\end{document}

%% file: authors.tex
\author{Itay M. Bloch}
\affiliation{Berkeley Center for Theoretical Physics, University of California, Berkeley, CA 94720, U.S.A.}
\affiliation{Theoretical Physics Group, Lawrence Berkeley National Laboratory, Berkeley, CA 94720, U.S.A.}

\author{Ana M. Botti}
\affiliation{\normalsize\it 
Fermi National Accelerator Laboratory, PO Box 500, Batavia IL, 60510, USA}
\affiliation{\normalsize\it Kavli Institute for Cosmological Physics, University of Chicago, Chicago, IL 60637, USA}

\author{Mariano Cababie}
\affiliation{\normalsize\it
Institut f\"ur Hochenergiephysik der \"Osterreichischen Akademie der Wissenschaften, 1050 Wien - Austria}
\affiliation{\normalsize\it
Atominstitut, Technische Universit\"at Wien, 1020 Wien - Austria}
\affiliation{\normalsize\it 
Fermi National Accelerator Laboratory, PO Box 500, Batavia IL, 60510, USA}

\author{Gustavo Cancelo}
\affiliation{\normalsize\it 
Fermi National Accelerator Laboratory, PO Box 500, Batavia IL, 60510, USA}

\author{Brenda A. Cervantes-Vergara}
\affiliation{\normalsize\it 
Universidad Nacional Aut\'onoma de M\'exico, Ciudad de M\'exico, M\'exico}

% Removed by JT Oct-23-2024
%\author{Michael Crisler}
%\affiliation{\normalsize\it 
%Fermi National Accelerator Laboratory, PO Box 500, Batavia IL, 60510, USA}

\author{Miguel Daal}
\affiliation{\normalsize\it 
 School of Physics and Astronomy, 
 Tel-Aviv University, Tel-Aviv 69978, Israel}

\author{Ansh Desai}
\affiliation{\normalsize\it 
Department of Physics and Institute for Fundamental Science, University of Oregon, Eugene, Oregon 97403, USA}

\author{Alex Drlica-Wagner}
\affiliation{\normalsize\it 
Fermi National Accelerator Laboratory, PO Box 500, Batavia IL, 60510, USA}
\affiliation{\normalsize\it Kavli Institute for Cosmological Physics, University of Chicago, Chicago, IL 60637, USA}
\affiliation{\normalsize\it  Department of Astronomy and Astrophysics, University of Chicago, Chicago IL 60637, USA}

 \author{Rouven Essig}
\affiliation{\normalsize\it 
C.N.~Yang Institute for Theoretical Physics, Stony Brook University, Stony Brook, NY 11794, USA}

 \author{Juan Estrada}
\affiliation{\normalsize\it 
Fermi National Accelerator Laboratory, PO Box 500, Batavia IL, 60510, USA}

\author{Erez Etzion}
\affiliation{\normalsize\it 
 School of Physics and Astronomy, 
 Tel-Aviv University, Tel-Aviv 69978, Israel}

\author{Guillermo Fernandez Moroni}
\affiliation{\normalsize\it 
Fermi National Accelerator Laboratory, PO Box 500, Batavia IL, 60510, USA}

\author{Stephen E. Holland}
\affiliation{\normalsize\it 
Lawrence Berkeley National Laboratory, One Cyclotron Road, Berkeley, California 94720, USA}

\author{Jonathan Kehat}
\affiliation{\normalsize\it 
 School of Physics and Astronomy, 
 Tel-Aviv University, Tel-Aviv 69978, Israel}

\author{Ian Lawson}
\affiliation{\normalsize\it SNOLAB, Lively, ON P3Y 1N2, Canada}

\author{Steffon Luoma}
\affiliation{\normalsize\it SNOLAB, Lively, ON P3Y 1N2, Canada}

 \author{Aviv Orly}
\affiliation{\normalsize\it 
 School of Physics and Astronomy, 
 Tel-Aviv University, Tel-Aviv 69978, Israel}

\author{Santiago E. Perez}
\affiliation{\normalsize\it 
Fermi National Accelerator Laboratory, PO Box 500, Batavia IL, 60510, USA}
\affiliation{\normalsize\it 
Universidad de Buenos Aires, Facultad de Ciencias Exactas y Naturales, Departamento de Física, Buenos Aires, Argentina}
\affiliation{\normalsize\it 
CONICET - Universidad de Buenos Aires, Instituto de Física de Buenos Aires (IFIBA). Buenos Aires, Argentina}

\author{Dario Rodrigues}
\affiliation{\normalsize\it 
Universidad de Buenos Aires, Facultad de Ciencias Exactas y Naturales, Departamento de Física, Buenos Aires, Argentina}
\affiliation{\normalsize\it 
CONICET - Universidad de Buenos Aires, Instituto de Física de Buenos Aires (IFIBA). Buenos Aires, Argentina}

\author{Nathan A. Saffold}
\affiliation{\normalsize\it 
Fermi National Accelerator Laboratory, PO Box 500, Batavia IL, 60510, USA}

\author{Silvia Scorza}
\affiliation{\normalsize\it 
Univ. Grenoble Alpes, CNRS, Grenoble INP, LPSC-IN2P3, Grenoble, 38000, France}

 \author{Miguel Sofo-Haro}
\affiliation{\normalsize\it 
Fermi National Accelerator Laboratory, PO Box 500, Batavia IL, 60510, USA}
\affiliation{Universidad Nacional de C\'ordoba, CNEA/CONICET, C\'ordoba, Argentina}

% Removed by JT Oct-23-2024
%\author{Leandro Stefanazzi}
%\affiliation{\normalsize\it 
%Fermi National Accelerator Laboratory, PO Box 500, Batavia IL, 60510, USA}

 \author{Kelly Stifter}
\affiliation{\normalsize\it 
Fermi National Accelerator Laboratory, PO Box 500, Batavia IL, 60510, USA}

\author{Javier Tiffenberg}
\affiliation{\normalsize\it 
Fermi National Accelerator Laboratory, PO Box 500, Batavia IL, 60510, USA}

\author{Sho Uemura}
\affiliation{\normalsize\it 
Fermi National Accelerator Laboratory, PO Box 500, Batavia IL, 60510, USA}

\author{Edgar Marrufo Villalpando}
\affiliation{\normalsize\it Kavli Institute for Cosmological Physics, University of Chicago, Chicago, IL 60637, USA}

\author{Tomer Volansky}
\affiliation{\normalsize\it 
 School of Physics and Astronomy,   
 Tel-Aviv University, Tel-Aviv 69978, Israel}

% Added by JT Oct-23-2024
\author{Federico Winkel}
\affiliation{\normalsize\it 
Universidad de Buenos Aires, Facultad de Ciencias Exactas y Naturales, Departamento de Física, Buenos Aires, Argentina}
\affiliation{\normalsize\it 
CONICET - Universidad de Buenos Aires, Instituto de Física de Buenos Aires (IFIBA). Buenos Aires, Argentina}

\author{Yikai Wu}
\affiliation{\normalsize\it 
C.N.~Yang Institute for Theoretical Physics, Stony Brook University, Stony Brook, NY 11794, USA}
\affiliation{\normalsize\it 
Department of Physics and Astronomy, Stony Brook University, Stony Brook, NY 11794, USA} 

\author{Tien-Tien Yu}
\affiliation{\normalsize\it 
Department of Physics and Institute for Fundamental Science, University of Oregon, Eugene, Oregon 97403, USA}

\collaboration{The SENSEI Collaboration }

% sho's notes:
% people who were borderline for 2023 paper (Silvia, Leo, Edgar)
% people who have left (Yaron, Kelly)
% people whose status I don't know (Prakruth, Mariano)